\documentclass[graybox]{svmult}
\usepackage{helvet}
\usepackage{courier}
\usepackage{makeidx}
\usepackage{graphicx}
\usepackage{multicol}
\usepackage{amsmath}
\usepackage{amssymb}
\usepackage{cite}

\makeindex
\begin{document}

\title*{Influence of Fractal Embedding in Three-Dimensional Euclidean
Space on Wave Propagation in Electro- Chromodynamics
}
\author{V.A. Okorokov and E.V. Sandrakova}
\institute{V.A. Okorokov \at NRNU MEPhI, Kashirskoe Shosse 31,
115409 Moscow, Russia \\ \email{VAOkorokov@mephi.ru} \and E.V.
Sandrakova \at NRNU MEPhI, Kashirskoe Shosse 31, 115409 Moscow,
Russia \\ \email{e.Sandrakova@yandex.ru}}
\maketitle

\abstract{In this paper two zero-dimensional compact sets with
equal topological and fractal dimensions but embedded in Euclidean
space by different ways are under study. Diffraction of plane
electromagnetic wave propagated and reflected by fractal surfaces
is considered for each of these compact sets placed in vacuum. It
is obtained, that the embedding of compact influences on
characteristics of wave in final state. Thus, the embedding of
Cantor set in Euclidean space is additional property of a fractal
which can be important both for applications of fractal
electrodynamics and for physics of strong interactions.}

\section{Geometry and strong interactions}
\label{sec:1} Geometry is believed to play a crucial role in the
formulation and understanding of mathematical methods of modern
physics theories. This statement concerns, in particular,
electrodynamics and quantum chromodynamics (QCD). Objects with
high irregular geometry possessing properties impossible to
describe in the framework of Euclid's geometry are wide spread in
nature and are encountered in many fields of science also.
Usually, these objects are fractals. It seems the search is
important for connections between geometrical characteristics of
these structures and physical quantities which describe its
(structures) properties with respect to electromagnetic and / or
strong interactions.

QCD as non-Abelian theory contains non-trivial topological gauge
field configurations, which are localized objects (instantons,
sphalerons) with average linear size $\sim 1/3$ fm in Euclidian
space. Study in the field of pion femtoscopy allows to get the
following numerical estimations for volume of matter created in
relativistic heavy  ion collisions: $V_{f} \sim 2.5 \times 10^{3}$
fm$^{3}$ at highest RHIC energy $\sqrt{s_{\mbox{\tiny NN}}} = 200$
GeV
\cite{Okorokov-ISHEPP2006,Abelev-PR-C80-024905-2009,Aamodt-PL-B696-328-2011}
and $V_{f} \sim 5.0 \times 10^{3}$ fm$^{3}$ at LHC energy
$\sqrt{s_{\mbox{\tiny NN}}} = 2.76$ TeV
\cite{Aamodt-PL-B696-328-2011}. In spite of these estimations
correspond to the freeze-out stage one can expect significant
amount of instantons / sphalerons can be inside such volume and
these topologically non-trivial objects can form clusters and
structures with highly irregular geometry. One can suggest at
qualitative level at least that such structures can be
characterized by fractional topology charge ($\mathcal{N}$). This
hypothesis does not contradict with general case of solutions on
solitons in quantum field theory \cite{Goldstone-PRL-47-986-1981}.
It was found in \cite{Goldstone-PRL-47-986-1981} that the fermion
number on kinks in one dimension or on magnetic monopoles in three
dimensions is, in general, a transcendental function of the
coupling constant of the gauge theories. On the one hand, as was
indicated in \cite{Goldstone-PRL-47-986-1981}, the direct utility
of results for fractional topology charge for high energy physics
seems problematical and it is more likely that corresponding
effects do arise in condensed matter systems. But on the other
hand it seems there are several types of objects (merons,
fractons) which are characterized by non-integer $\mathcal{N}$ and
play an important role for explanation of structure of the QCD
vacuum and such striking phenomenon as confinement of color
charge. Although instanton solutions with $\mathcal{N}=1$ provide
an successful understanding of chiral symmetry breaking, in the
dilute gas \cite{Callan-PRD-18-4684-1978} and instanton liquid
approximations \cite{Shuryak-LectureNote-71-2004}, they do not
lead to confinement: it was shown analytically in
\cite{Callan-PRD-18-4684-1978} that contribution of the dilute
instanton gas to the potential between two static color charges
does not to produce confinement. The hypothetical objects with
topological charge 1/2, founded in
\cite{deAlfrano-PLB-65-163-1976} and called merons, are more
strongly disordering objects than instantons. The merons are
objects into which instantons can dissociate at large sizes, when
the coupling and quantum fluctuations get big enough. Unlike
instantons, which interact at large distances weakly, as dipole,
and cannot effectively disorder the Wilson loop, the merons
interact like charge and can do so
\cite{Shuryak-LectureNote-71-2004}. Although a free meron has an
action $S$ that diverges logarithmically with the volume of the
space-time region $\mathcal{V}$, a pair of merons has a finite $S$
that depends on $\ln\,r$, where $r$ is the Euclidean separation.
If the coupling is strong enough that the logarithmic potential is
weaker than the entropy associated with the $r$ ($\sim r^{3}$),
meron pairs could dissociate and form a meron gas
\cite{Lenz-AnnP-323-1536-2008}. Since meron fields only fall off
as $\sim 1/x$, they are expected to be more effective in
disordering than singular gauge instantons and, moreover, the
ensemble of these pseudoparticles, and thus potentially be a more
effective mechanism for confinement
\cite{Lenz-AnnP-323-1536-2008}. Thus merons were proposed as one
of the possible, at least, mechanisms for confinement
\cite{Callan-PLB-66-375-1977,Callan-PRD-17-2717-1978,Callan-PRD-19-1826-1979}.
Fractons, also solutions of the Yang--Mills equations of motion
with fractional values of $\mathcal{N}$, appear on the
four-dimensional torus, $\mathbf{T}^{4}$, when twisted boundary
conditions are imposed \cite{tHooft-NPB-153-141-1979}. The
possible relevance of these objects to confinement phenomenon was
pointed out in \cite{tHooft-CMP-81-267-1981}, and a scenario for
confinement based on the fractional topological charge was
proposed in
\cite{Garcia-PLB-235-117-1990,Garcia-JPA-26-2667-1993,Gonzalez-NPB-459-337-1996}.
The deep interrelation between key features of $\mbox{SU}(2)$
Yang--Mills theory, particular, confinement and objects with
fractional values of $\mathcal{N}$, namely, merons is discussed in
more detail in
\cite{Lenz-AnnP-323-1536-2008,Steele-PRL-85-4207-2000,Montero-PLB-533-322-2002,Lenz-PRD-69-074009-2004,
Negele-NPProcSuppl-140-629-2005}. Therefore the objects or / and
structures with non-integer $\mathcal{N}$ can be important for
understanding of structure of the QCD vacuum and, as consequence,
wide set of nonpeturbative phenomena such as confinement, hadron
structure etc. Moreover the intensive investigations on RHIC and
LHC demonstrate that the final state matter created in
relativistic heavy ion collisions in energy domain, at least,
$\sqrt{s_{\mbox{\tiny NN}}} \simeq 0.04 - 2.76$ TeV is quark-gluon
system characterized by very low viscosity and strong coupling,
i.e. (quasi)ideal quark-gluon liquid -- so called strongly coupled
quark-gluon plasma (sQGP). The theoretical and experimental
studies indicate that this matter will remains strongly coupled
system up to highest experimentally available energies. The new
corresponding physics realm called "QCD of condensed matter" is
developed rapidly at the present time
\cite{Okorokov-YaF-72-147-2009}. Thus the objects with fractional
topological charge can be some interesting in point of view both
the problem of confinement of color charges in QCD and for study
of strongly coupled systems (sQGP). Suggestion with respect to
production of large amount of topologically non-trivial objects is
confirmed by estimations which show creation about 400
sphaleron-type clusters in $\mbox{Au+Au}$ collisions at
$\sqrt{s_{\mbox{\tiny NN}}} = 200$ GeV
\cite{Shuryak-NP-A715-289c-2003}. Hypothesis concerned on complex
and highly irregular geometry of QCD vacuum in Euclidean space
does not contradict both lattice calculations
\cite{Ilgenfritz-NPPS-B153-328-2006,Weinberg-XXIV-2006} and
fractal-like approximation of structures in distributions of
topological charge density \cite{Horvath-PR-D68-114505-2003}. The
picture that has emerged on basis of these investigations is quite
different from earlier conceptions like the dilute instanton gas
or even the instanton liquid \cite{Schierholz-ICHEPP2006}.

Structures with non-trivial topology in QCD vacuum are believed to
relate with some nonperturbative phenomena like large mass of
$\eta'$ meson, the axial $U(1)$ anomaly, violation of some
fundamental symmetries ($\mathcal{P}/\,\mathcal{CP}$) in hot QCD
matter. The important note is that the coupling of these
non-trivial topological field configurations to electromagnetic
field induced by the axial anomaly in the framework of
Maxwell--Chern--Simons electrodynamics can lead to the
experimentally observable effects in the presence of external
strong magnetic field (so called "chiral magnetic effect")
\cite{Kharzeev-AnnPhys-325-205-2010}. At present it is obtained
that extremely intense (electro)magnetic field is generated in
relativistic heavy ion collisions
\cite{Kharzeev-NP-A803-227-2008,Okorokov-arXiv-0908.2522}. The
evidence for the chiral magnetic effect has been found recently
both from numerical lattice QCD calculations
\cite{Buividovich-PR-D80-054503-2009} and from several experiments
at RHIC
\cite{Abelev-PRL-103-251601-2009,Abelev-PR-C81-054908-2010} and
LHC \cite{Abelev-PRL-110-012301-2013} for the field of strong
interaction physics. Therefore the study seems important for
interaction of electromagnetic field with topologically
non-trivial structures which can be characterized as fractal-like
ones.

\section{Wave propagation in various fractals}
\label{sec:2} In this paper the problem of electromagnetic wave
propagation in fractal environment is considered. At present it is
unknown what structures namely can be formed by topologically
non-trivial QCD-objects, i.e. type of possible fractals is unknown
exactly. Thus one can makes only some hypothesis concerning the
type of possible highly irregural structures or consider some
examples based on known fractals. It is supposed that two
zero-dimensional compact sets $\mathbf{A}$ and $\mathbf{C}$ with
different topological properties and equal fractal dimensions,
i.e.
\begin{eqnarray}
\left. \begin{array}{l}
\mbox{dim}\mathbf{A}=0, \mbox{dim}\mathbf{C}=0, \\
\mbox{dim}_{\mbox{\scriptsize
H}}\mathbf{A}=\mbox{dim}_{\mbox{\scriptsize H}}\mathbf{C},
\end{array}\right\}\label{eq:01}
\end{eqnarray}
are placed in vacuum. Here $\mbox{dim}$ -- topological dimension,
$\mbox{dim}_{\mbox{\scriptsize H}}$ -- Hausdorff dimension.

\begin{definition}
A zero-dimensional compact set $\mathbf{K}$ is named \textit{wild}
or \textit{wild embedded} in three-dimensional Euclidean space
$\mathbf{E}^{3}$ if there is no homeomorphism of the space
$\mathbf{E}^{3}$ on itself at which the compact $\mathbf{K}$ would
be transformed in zero-dimensional compact lied on a straight line
$\mathbf{E}^{1} \subset \mathbf{E}^{3}$. In the opposite case the
compact $\mathbf{K}$ is named \textit{tame}.\label{def:01}
\end{definition}

In the framework of group theory this definition can be
reformulated by the following way: a zero-dimensional compact set
is defined as \textit{wild} in three-dimensional Euclidean space
$\mathbf{E}^{3}$ if its complement in $\mathbf{E}^{3}$ has non
trivial fundamental group, otherwise the zero-dimensional compact
set is \textit{tame} in $\mathbf{E}^{3}$.

Lets compact set $\mathbf{A}$ is the zero-dimensional Antoine
compact set \cite{Antoine-JMPA-4-221-1921}. The Antoine compact
set can be represented by the follow way:
\begin{equation}
\mathbf{A}=\bigcap\limits_{k=1}^{+\infty}\bigcup\limits_{i=1}^{3^{k}}\mathbf{T}_{ki},
\label{eq:02}
\end{equation}
where $\mathbf{T}_{ki}$ is the polyhedral complete torus for which
the relations are valid $\forall \,
i=1,2,\dotsc,3^{k},~k=0,1,\dotsc:~\mathbf{T}_{ki} \cap
\mathbf{T}_{kj}=\varnothing$ at $i \ne j$;
$\mbox{diam}\mathbf{T}_{ki} < (k+1)^{-1}$; the three complete tori
of next "level" $\mathbf{T}_{k+1i_{j}},~j=1,2,3$ are inside of
each complete torus $\mathbf{T}_{ki}$ with that the each of
complete tori $\mathbf{T}_{k+1i_{j}}$ is locking with two another
ones and chain of complete tori
$\bigcup_{j=1}^{3}\mathbf{T}_{k+1i_{j}}$ can not be drawn together
into point in $\mathbf{T}_{ki}$. It should be noted that the chain
of complete tori $\mathbf{T}_{11} \cup \mathbf{T}_{12} \cup
\mathbf{T}_{13}$ lies in the complete torus $\mathbf{T}_{0}
\subset \mathbf{E}^{3},~\mbox{diam}\mathbf{T}_{0} < 1$. The
Antoine compact set $\mathbf{A}$ is wild compact one
\cite{Antoine-JMPA-4-221-1921}.

The zero-dimensional compact set $\mathbf{C}$, homeomorphous to
the compact set $\mathbf{A}$, was constructed in
\cite{Sandrakova-MC-85-98-1971} as following:
\begin{equation}
\mathbf{C}=\bigcap\limits_{k=0}^{+\infty}\mathbf{C}_{k}=\bigcap\limits_{k=0}^{+\infty}
\bigcup\limits_{j=1}^{3^{k}}\mathbf{U}_{kj}.\label{eq:03}
\end{equation}
Here $\mathbf{C}_{0}=\mathbf{T}'_{0}$ is the complete torus in
$\mathbf{E}^{3}$ and $\mbox{diam}\mathbf{T}'_{0} < 1$. Following
relations are valid: $\mathbf{U}_{kj}=\mathbf{T}'_{kj}$, $j \ne
3n$; $\mathbf{U}_{kj}=\mathbf{Q}'_{kj}, j=3n$; $\mathbf{U}_{ki}
\cup \mathbf{U}_{kj}=\varnothing$, $i \ne j$; and
$\left(\bigcup_{j=3m+1}^{3m+3}\mathbf{U}_{kj}\right) \subset
\mbox{int}\mathbf{U}_{k-1l}$ for arbitrary $l$, where
$\mathbf{T}'_{kj}$ -- polyhedral complete torus with
$\mbox{diam}\mathbf{T}'_{k+1_{j}} \! < \! (k+2)^{-1}$,
$\mathbf{Q}'_{kj}$ is the polyhedral three-di\-men\-sio\-nal cell
with $\mbox{diam}\mathbf{Q}'_{k+1_{j}}\! < \! (k+2)^{-1}$. At
construction of the $\mathbf{C}_{k+1}$ two complete tori
$\mathbf{T}'_{k+1i_{1}},~\mathbf{T}'_{k+1i_{2}}$ locked each to
other and without intersection and cell $\mathbf{Q}'_{k+1i_{3}}$
is constructed inside of each $\mathbf{U}_{ki}$ so as to
$\left(\bigcup_{j=1}^{2}\mathbf{T}'_{k+1i_{j}}\right) \bigcap
\mathbf{Q}'_{k+1i_{3}}=\varnothing$. Thus
$\mathbf{C}_{k+1}=\bigcup_{i=1}^{3^{k+1}}\mathbf{U}_{k+1i}$ i.e.
$\mathbf{C}_{k+1}$ is the unification of the pairs of polyhedral
complete tori without intersections
$\mathbf{T}'_{k+1i_{j}},~j=1,2$ and polyhedral three-dimensional
cells $\mathbf{Q}'_{k+1i_{3}}$ with that
$\mbox{diam}\mathbf{U}_{k+1i} \to 0$ together with $(k+2)^{-1}$ at
$k \to \infty$. The object $\mathbf{C}$ defined by (\ref{eq:03})
is compact set because each of its elements $\mathbf{C}_{k}$ is
compact set. By the construction $\mathbf{C} \subset
\bigcup_{i=1}^{3^{k}}\mbox{int}\mathbf{U}_{ki}$ for each $k$. It
follows that $\mathbf{C}$ is the sum of pairs non-intersected
openly-closed sets: $\mathbf{C} \cap \mbox{int}\mathbf{U}_{k}$,
which diameters are tend to zero with increasing of $k$.
Therefore, one can obtain $\mbox{diam}\mathbf{C}=0$. In
\cite{Sandrakova-MC-85-98-1971} it was proved that compact set
$\mathbf{C}$ is the tame compact set in $\mathbf{E}^{3}$.

It seems the circumstance is very important that both the wild
Antoine compact set $\mathbf{A}$ and the tame compact $\mathbf{C}$
are fractals. Compact sets $\mathbf{A}$ and $\mathbf{C}$ are
homeomorphous with the Cantor perfect set $\mathbf{\tilde K}$ lied
on a straight line $\mathbf{E}^{1} \subset \mathbf{E}^{3}$.

\begin{definition}
Psedoisotopy $\mathcal{F}_{t}$ of the space $\mathbf{E}^{3}$ is
named homotopy $\mathcal{F}_{t}, t \in [0,1],~\mathcal{F}_{t}:
\mathbf{E}^{3} \to \mathbf{E}^{3}$ such as to $\mathbf{F}_{t}$ is
homeomorphous mapping of the space $\mathbf{E}^{3}$ on itself for
$t \in [0,1)$ but at $t=1$ the $\mathcal{F}_{1}$ is continuous
mapping of the $\mathbf{E}^{3}$ on itself. If
$\mathcal{F}_{0}=\mathcal{I}$, where $\mathcal{I}$ is the identity
map, then one can speaks that $\mathcal{F}_{t}$ is pseudoisotopy
from identity map.\label{def:02}
\end{definition}

In \cite{Sandrakova-MC-85-98-1971} it was shown, that for the
compact set (\ref{eq:03}) such pseudoisotopy can be constructed
$\mathcal{F}_{t}, t \in [0,1],~\mathcal{F}_{0}=\mathcal{I},
\mathcal{F}_{t}: \mathbf{E}^{3} \to \mathbf{E}^{3}$, which
transforms the tame compact set (\ref{eq:03}) in the wild compact
set (\ref{eq:02}). In \cite{Bing-PJM-11-435-1961} it was shown,
that the set of points of wildness for zero-dimensional compact
set $\mathbf{K}$ in $\mathbf{E}^{3}$ is either non-counting or
empty. In \cite{Crownover-book1-1999,Okorokov-book1-2005} the
following theorem was proved.

\begin{theorem}
Lets $\mathbf{K}_{1}$ and $\mathbf{K}_{2}$ are compact sets in
$\mathbf{E}^{3}$ and lets $\bf{f}: \mathbf{K}_{1} \to
\mathbf{K}_{2}$, where $\bf{f}$ is homeomorphism. Moreover
$\bf{f}$ and $\bf{f}^{-1}$ satisfy the Lipshitz condition. Then
$\mathbf{K}_{1}$ and $\mathbf{K}_{2}$ have equal fractal
dimensions.
\end{theorem}

As homeomorphism $\bf{f}: \mathbf{K}_{1} \to \mathbf{K}_{2}$ one
can takes the mapping $\mathcal{F}_{1}$ constructed by the
pseudoisotopy $\mathcal{F}_{t}, t \in [0,1],~\mathcal{F}_{t}:
\mathbf{E}^{3} \to \mathbf{E}^{3}$ so as to it (i.e. this map)
satisfies the Lipshitz condition \cite{Sandrakova-MC-85-98-1971}.
It follows that $\mbox{dim}_{\mbox{\scriptsize
H}}\mathbf{A}=\mbox{dim}_{\mbox{\scriptsize H}}\mathbf{C}$, i.e.
compact sets (\ref{eq:02}) and (\ref{eq:03}) are fractals with
equal fractal dimension $D$.

Lets the complete torus $\mathbf{T}_{0}$ is filled by some matter
with refraction index $n_{1}$ differs from the refraction index of
environment (vacuum) $n_{0}$. The right Cartesian coordinates are
chosen as the coordinate system with axis OZ directed
horizontally, with origin of coordinates coincided with the center
of symmetry of complete torus $\mathbf{T}_{0}$, and with the plane
OXY included the mean line of the complete torus under
consideration. Lets the plane harmonic wave $\varPsi(x,y,z)$ falls
on the $\mathbf{T}_{0}$ with that the wave length satisfies the
relation $\lambda \ll \mbox{diam}\mathbf{T}_{0}$.

At the construction step with arbitrary number $k$ for Antoine
compact set $\mathbf{A}$ one considers the chain of complete tori
of corresponding level $\bigcup_{j=1}^{3}\mathbf{T}_{ki_{j}}$
inside some fixed, for example, $i$-th complete torus
$\mathbf{T}_{ki}$ of previous level. Note that the chain
$\bigcup_{j=1}^{3}\mathbf{T}_{ki_{j}}$ is not drawn together into
point in $\mathbf{T}_{ki}$. The wave propagated in
$\mathbf{T}_{ki}$ and passed through complete torus
$\mathbf{T}_{ki_{j}}~(j=1,2,3)$, collides with obstacle in the
places of locking $\mathbf{T}_{ki_{1}}$ with $\mathbf{T}_{ki_{2}}$
and $\mathbf{T}_{ki_{3}}$. Thus the wave diffraction is formed on
the each step with number up to some $\tilde{k}$, where
$\tilde{k}$ is defined by the relation
$\mbox{diam}\mathbf{T}_{\tilde{k}\dotsc1} \! < \!
(\tilde{k}+1)^{-1} \! < \! \lambda$, which means the wave can not
distinguishes complete tori with diameters smaller than wave
length.

In the case of $\mathbf{C}$ the compact set
$\mathbf{U}_{kj}=\mathbf{Q}'_{kj}$ at $j=3n$ is the polyhedral
three-dimensional cell which creates splits with complete tori
$\mathbf{U}_{kj}=\mathbf{T}_{kj}$ at $j \ne 3n$. Because $\forall
\, j: \mathbf{U}_{kj}$ are polyhedrons for the compact set
$\mathbf{C}$ the construction can be made so that the width of
appeared split will be smaller than up to some step number
$\tilde{k}$. Thus diffraction from split will be generated in this
case instead of diffraction from obstacle.

\section{Summary}\label{sec:03}

The main aim of this paper is qualitative study of influence of
fractal embedding on physics characteristics of wave propagation.
It seems that the relations between geometrical (embedding)
properties of some fractal object and physical properties of waves
interacted with this object can be important both for fundamental
studies and for applications.

Therefore, based on the qualitative consideration the following
conclusion can be made. In spite of (\ref{eq:01}), embedding way
of the wild compact set $\mathbf{A}$ and tame compact set
$\mathbf{C}$ in $\mathbf{E}^{3}$ can influences on physics
characteristics of wave and on distribution of intensity, i.e.
embedding is the one more characteristic of fractal which can be
important for applications of fractal geometry in physics. There
are various experiments for propagation of electromagnetic waves
in fractal environments characterized by different Hausdorff
dimensions (see, for example, \cite{Potapov-book1-2005}). But
unfortunately there is no experimental study of influence of
fractal embedding on physics characteristics of wave propagation.
This paper is the first qualitative study of the problem.
Therefore future investigations are necessary for derivation of
numerical estimations and verification of this hypothesis at
quantitative level.

\end{document}